# Believe It or Not:
# Adding Belief Annotations to Databases

Wolfgang Gatterbauer, Magdalena Balazinska, Nodira Khoussainova, and Dan Suciu
Department of Computer Science and Engineering,
University of Washington, Seattle, WA, USA
{gatter, magda, nodira, suciu}@cs.washington.edu

## ABSTRACT

We propose a database model that allows users to annotate data with *belief statements*. Our motivation comes from scientific database applications where a community of users is working together to assemble, revise, and curate a shared data repository. As the community accumulates knowledge and the database content evolves over time, it may contain *conflicting information* and members can *disagree* on the information it should store. For example, Alice may believe that a tuple should be in the database, whereas Bob disagrees. He may also insert the reason why he thinks Alice believes the tuple should be in the database, and explain what he thinks the correct tuple should be instead.

We propose a formal model for *Belief Databases* that interprets users' annotations as belief statements. These annotations can refer both to the base data and to other annotations. We give a formal semantics based on a fragment of multi-agent epistemic logic and define a query language over belief databases. We then prove a key technical result, stating that every belief database can be encoded as a *canonical Kripke structure*. We use this structure to describe a relational representation of belief databases, and give an algorithm for translating queries over the belief database into standard relational queries. Finally, we report early experimental results with our prototype implementation on synthetic data.

## 1. INTRODUCTION

In many sciences today, a *community of users* is working together to assemble, revise, and curate a shared data repository. Examples of such collaborations include identifying functions of particular regions of genetic sequences [39], curating databases of protein functions [11, 46], identifying astronomical phenomena on images [43], and mapping the diversity of species [37].

As the community accumulates knowledge and the database content evolves over time, it may contain *conflicting information* and members may *disagree* on the information it should store. Relational database management systems (DBMSs) today can help these communities manage their shared data, but provide limited support for managing conflicting facts and conflicting opinions about the correctness of the stored data.

The recent concept of *database annotations* aims to address this need: annotations are commonly seen as superimposed information that helps to explain, correct, or refute base information [36] without actually changing it. Annotations have been recognized by scientists as an essential feature for new generation database management systems [4, 8, 19], and efficient management of annotations has become the focus of much recent work in the database community [7, 11, 13, 15, 23, 24]. Still, the semantic distinction between base information and annotations remains blurred [10]. Annotations are simply additional metadata added to existing data [44] without unique and distinctive semantics.

In discussions with scientists from forestry and bioengineering, we have seen the need for an *annotation semantics* that helps collaborating community members engage in a *structured discussion* on both content and each other's annotations: scientists do not only want to insert their own annotations but also want to be able to respond to other scientists' annotations. Such annotation semantics creates several challenges for a database system. First, it needs to allow for *conflicting annotations:* Users should be able to use annotations to indicate conflicts between what they believe and what others believe. The database should allow and expose those conflicts. Second, it should also support *higher-order annotations*. Users should be able to annotate not only content but also other users' annotations. And, finally, the additional functionality should be supported *on top of* a standard DBMS with a simple extension of SQL. Any new annotation model should take advantage of existing state-of-the art in query processing.

To address these challenges, we introduce the concept of a *belief database*. A belief database contains base information in the form of ground tuples, annotated with belief statements. It represents a set of different *belief worlds*, each one for one type of belief annotation, i.e. the beliefs of a particular user on ground tuples, or on another user's beliefs. These belief worlds follow an *open world assumption* and may be overlapping and partially conflicting with each other. The formal semantics of belief annotations is defined in terms of multi-agent epistemic logic [21]. This semantics can be represented by an appropriate *canonical Kripke structure* which, in turn, can be represented in the standard relational model and, hence, on top of a standard RDBMS. We also introduce *belief conjunctive queries*, a simple, yet versatile query language that serves as interface to a belief database and consists of conjunctive queries with belief assertions. In addition to retrieving facts believed or not believed by certain users, this language can also be used to query for agreements or disagreements between users. We describe an algorithm for translating belief conjunctive queries into non-recursive Datalog (and, hence, to SQL). We have implemented a prototype *Belief Database Management System* (BDMS), and describe a set of preliminary experiments validating the feasibility of translating belief queries into SQL.

The structure of this paper follows its contributions:
- We describe a motivating application, and give examples and a syntax for BeliefSQL (Sect. 2).
- We define a data model and a query language for belief databases (Sect. 3).
- We describe the canonical Kripke structure that enables implementing belief databases (Sect. 4).
- We describe a relational representation of belief databases and the translation of queries and updates over this canonical representation (Sect. 5).
- We validate our model and report on experiments with our prototype BDMS (Sect. 6).

The paper ends with an overview of related work (Sect. 7) and conclusions (Sect. 8).

## 2. MOTIVATING APPLICATION

In this section, we present a motivating application that we use as running example throughout this paper. The scenario is based on the NatureMapping project whose goal is to record biodiversity of species in the US state of Washington [37]. Participating community members volunteer to submit records of animal sightings from the field. Each observation includes user-id, date, location, species name, and various options to comment on the observation, such as details about how the animal was identified (e.g., animal tracks were found). As sightings are reported by non-experts, they can contain errors. In fact, even experts sometimes disagree on the exact species of a sighted animal.

In the current protocol, a single expert in forestry (the principal investigator) manually curates all the entries before inserting them into the database, which results in significant delays and does not allow the application to scale to a larger number of volunteers. In this setting, a *Belief Database Management System* (BDMS) can address this challenge by allowing multiple experts to annotate, thus streamlining the curation process. Graduate students, technicians, and expert users can all contribute their beliefs to annotate the data, thus proving a *collaborative curation process*. They can, for example, disagree with individual sightings, if in their judgment the sighting is incorrect, and annotate the data accordingly. They can also correct a sighting by annotating it with corrected values they believe more plausible than those provided by the volunteers in the field. And they can also suggest *explanations* for other users' annotations, thus leading to *higher-order annotations*.

```
select    selectlist
from      (((BELIEF user)+ not?)? relationname)+
where     conditionlist

insert    into ((BELIEF user)+ not?)? relationname
          values

delete    from ((BELIEF user)+ not?)? relationname
where     conditionlist

update    ((BELIEF user)+ not?)? relationname
set       value assignments
where     conditionlist
```

**Figure 1: Syntax of query and data manipulation commands in BeliefSQL.**

We now illustrate the use of a BDMS. We assume three users (Alice, Bob, and Carol) and a simplified database schema consisting of three relations:

Sightings(<u>sid</u>, uid, species, date, location)
Comments(<u>cid</u>, comment, sid)
Users(<u>uid</u>, name)

We refer to this schema as *external schema* since it presents the way users enter and retrieve data. Beliefs, in contrast, are stored transparently from users and can be manipulated via natural extensions to standard SQL (Fig. 1). We illustrate its usage through examples next.

Little Carol sees a bald eagle during her school trip and reports her sighting with the following insert:

$i_1$:insert into Sightings
    values ('s1','Carol','bald eagle','6-14-08','Lake Forest')

Bob, a graduate student, however, does not believe that Carol saw a bald eagle:

$i_2$:insert into BELIEF 'Bob' not Sightings
    values ('s1','Carol','bald eagle','6-14-08','Lake Forest')

Additionally, Bob does not believe that Carol could have seen a fish eagle, which looks similar to a bald eagle:

$i_3$:insert into BELIEF 'Bob' not Sightings
    values ('s1','Carol','fish eagle','6-14-08','Lake Forest')

This ensures that Bob still disagrees even if Carol's tuple is updated to species='fish eagle'. In both cases, Bob uses the external key 's1' to refer to the tuple with which he disagrees.

Alice, a field technician, believes there was a crow at Lake Placid because she found some black feathers. She does not insert a regular tuple as Carol did, but inserts only her own belief:

$i_4$:insert into BELIEF 'Alice' Sightings
    values ('s2','Alice', 'crow','6-14-08','Lake Placid')
$i_5$:insert into BELIEF 'Alice' Comments
    values ('c1','found feathers','s2')

Bob believes there cannot be any crows in the Lake Placid area. He wants to annotate the data with the following belief statements: (i) Bob believes that Alice saw a raven, not a crow; (ii) Bob believes that Alice believed that the feathers she found were black; and (iii) Bob believes the feathers were actually purple-black, suggesting they come from a raven, not a crow. The second and third belief statements above are Bob's suggestion why Alice may have made a mistake. These annotations are inserted into the BDMS as follows:

$i_6$:insert into BELIEF 'Bob' Sightings
  values ('s2','Alice','raven','6-14-08','Lake Placid')
$i_7$:insert into BELIEF 'Bob' BELIEF 'Alice' Comments
  values ('c2','black feathers','s2')
$i_8$:insert into BELIEF 'Bob' Comments
  values ('c2','purple-black feathers','s2')

Notice here the important role of the higher-order belief statement: "Bob believes that Alice believes that the feathers were black"; this is how Bob explains his disagreement with Alice. Such explanations are quite common in a collaborative data curation process, and it is important for a BDMS to support them.

At this point we have recorded eight belief statements in the database. In the following section, we adopt the formalism of *multi-modal logic* [25] and write $\Box_u t^+$ for the assertion "user $u$ believes tuple $t$". Figure 2 illustrates with our eight statements. Note that in practice, a BDMS needs to keep additional information in its *internal schema*, which we describe in Sect. 5.

Finally, we illustrate two queries over the belief database. The first query asks for sightings at Lake Forest believed by Bob. It returns ('s2','Alice','raven'):

$q_1$: select  S.skey, S.uid, S.species
   from    Users as U, BELIEF U.uid Sightings as S
   where   U.name = 'Bob'
   and     S.location = 'Lake Forest'

The second query retrieves entries on which users disagree with what Alice believes:

$q_2$: select  U2.name, S1.species, S2.species
   from    Users as U1, Users as U2,
           BELIEF U1.uid Sightings as S1,
           BELIEF U2.uid Sightings as S2,
   where   U1.name = 'Alice'
   and     S1.sid = S2.sid
   and     S1.species <> S2.species

The BDMS returns ('Bob','crow','raven'), implying that Bob disagrees with Alice's crow sighting.

## 3. FORMAL SETUP

We introduce here the basic notion of a *belief database*, which enriches a standard database with annotations of users' beliefs. Informally, a belief database represents a set of incomplete and consistent database instances. Depending on which tuples they share or do not share, any two such instances can be mutually disjoint, overlapping, contained or partly conflicting.

**Standard relational background.** We fix a relational schema $\mathcal{R} = (R_1, \ldots R_r)$ and assume that each

Ground tuples without annotations
Sightings

| | sid | uid | species | date | location |
|---|---|---|---|---|---|
| $s1_1$ | s1 | Carol | bald eagle | 6-14-08 | L.Forest |
| $s1_2$ | s1 | Carol | fish eagle | 6-14-08 | L.Forest |
| $s2_1$ | s2 | Alice | crow | 6-14-08 | L.Placid |
| $s2_2$ | s2 | Alice | raven | 6-14-08 | L.Placid |

Comments

| | cid | comment | sid |
|---|---|---|---|
| $c1_1$ | c1 | found feathers | s2 |
| $c2_1$ | c2 | black feathers | s2 |
| $c2_2$ | c2 | purple black feathers | s2 |

Belief annotations of ground tuples

$i_1$: $s1_1^+$
$i_2$: $\Box_{Bob}\, s1_1^-$
$i_3$: $\Box_{Bob}\, s1_2^-$
$i_4$: $\Box_{Alice}\, s2_1^+$
$i_5$: $\Box_{Alice}\, c1_1^+$
$i_6$: $\Box_{Bob}\, s2_2^+$
$i_7$: $\Box_{Bob}\Box_{Alice}\, c2_1^+$
$i_8$: $\Box_{Bob}\, c2_2^+$

**Figure 2:** Our running example. Left: Ground tuples inserted and annotated by different users. Conflicting tuples (like the crow and raven tuples) share the same *external key*. Internal keys (like $s1_1$ and $s1_2$) uniquely identify each tuple. Right: Belief annotations over the ground tuples written in the notation of multi-modal logic.

relation $R_i(att_{i1}, \ldots att_{il_i})$ with arity $l_i$ has a distinguished primary key $att_{i1}$, for which we alternatively write $key_i$ to make the key attribute explicit. In the context of belief databases, we call $\mathcal{R}$ the *external schema*, as this is how users see the non-annotated data, and denote $I$ a *conventional database instance* without annotations. An *incomplete database* is a set of conventional database instances $\{I_1, I_2, \ldots\}$ over a fixed schema $\mathcal{R}$ [32, 28]. For each relation $R_i$, denote $Tup_i$ the set of typed atomic tuples of the form $R_i(a_1, \ldots a_{l_i})$. Further denote $Tup = \bigcup_i Tup_i$ the domain of all tuples or the *tuple universe* of the schema. We further require that $Tup_i \cap Tup_j = \emptyset$ where $i \neq j$, i.e., each tuple $t \in Tup$ is uniquely associated with one relation of the schema. If $t \in Tup$ then $key(t)$ represents the typed value of the key attribute in $t$. Using this notation, *consistency* and conventional key constraints are defined as follows:

DEFINITION 1 (CONSISTENCY). *A database instance $I$ over a relation $R$ is* consistent *iff it satisfies the key constraints $\Gamma(I)$, i.e. no two tuples from the same relation share the same key:*

$\Gamma(I) \equiv \forall i. \forall t, t' \in Tup_i. (t, t' \in I \land t \neq t' \Rightarrow key(t) \neq key(t'))$

### 3.1 Belief worlds

A *belief world* is a set of positive and negative beliefs of a user about the database content or other user's beliefs, and represents a set of consistent database instances. For example, one belief world is what Alice believes, another one is what Bob believes Alice believes.

Negative beliefs arise naturally when users disagree about a ground fact or belief but do not have an alternative suggestion. In order to allow for such explicit negative database entries, the default has to consider a tuple *possible* before it is inserted as either positive or negative. This default corresponds to the *Open World Assumption* (OWA), and differs from conventional databases where every tuple that is not in the database is considered negated according to the *Closed World Assumption* (CWA) [40].

We next give a precise definition and semantics to a belief world based on incomplete databases:

DEFINITION 2 (BELIEF WORLD). *A belief world is a pair $W = (I^+, I^-)$, where both $I^+$ and $I^-$ are conventional database instances over the schema $\mathcal{R}$ that are, a priori, not required to satisfy the key constraints.*

DEFINITION 3 (SEMANTICS OF A BELIEF WORLD). *The semantics of a belief world $W = (I^+, I^-)$ is the incomplete database of instances $I$ over the schema $\mathcal{R}$ that contain all tuples from $I^+$, contain no tuples from $I^-$, and that satisfy the key constraints.*

$$[\![W]\!] = \{I \mid I^+ \subseteq I,\ I \cap I^- = \emptyset,\ \Gamma(I)\}$$

DEFINITION 4 (CONSISTENCY OF A BELIEF WORLD). *A belief world $W$ is* consistent *iff $[\![W]\!] \neq \emptyset$.*

PROPOSITION 5 (CONSISTENCY OF A BELIEF WORLD). *A belief world $W$ is consistent iff it satisfies the following two constraints:*

$$\Gamma_1(W) \equiv \Gamma(I^+)$$
$$\Gamma_2(W) \equiv \forall t \in I^+ : t \notin I^-$$

The above definitions and proposition state that a belief world is represented by two different database instances. Those have to fulfill two constraints in order to represent a consistent set of beliefs: $\Gamma_1$ is a standard key constraint on $I^+$, and $\Gamma_2$ requires that $I^+ \cap I^- = \emptyset$.

It is convenient to represent a belief world by combining the two instances $I^+$ and $I^-$ into a single table where each tuple has an additional *sign* attribute $s$ whose value is '+' for the tuples in $I^+$ and '−' for those in $I^-$. Figure 3 illustrates this with the belief world "Bob believes" from our running example. His version of Sightings has one positive and two negative records. For example, Bob believes that Alice saw a raven (tuple with sid = 's2'), but he does not believe that Carol saw a 'bald eagle' nor a 'fish eagle' (both tuples share sid ='s1', hence refer to the same sighting). This example illustrates why $I^-$ does not have to satisfy the key constraints: we want to allow a user to disagree with more than one alternative. This is needed, for example, if Alice adds a belief statement $i_9$ with the species 'fish eagle' as alternative explanation of Carol's entry $i_1$:

$i_1$ : ('s1','Carol','bald eagle','6-14-08','Lake Forest')$^+$

$i_9$ : $\Box_{\text{Alice}}$('s1','Carol','fish eagle','6-14-08','Lake Forest')$^+$

Here, $i_1$ and $i_9$ represent conflicting positive statements. But, in addition, Bob disagrees with both.

We now define positive and negative beliefs formally. Note that they correspond exactly to the concepts of *certain* and *impossible* tuples in incomplete databases.

DEFINITION 6 (POSITIVE AND NEGATIVE BELIEFS). *Let $W$ be a belief world. We say that a tuple $t$ is a* positive belief *for $W$ iff $t$ belongs to all instances in $[\![W]\!]$ and write $W \models t^+$. We say that a tuple $t$ is a* negative belief *for $W$ iff $t$ belongs to no instance in $[\![W]\!]$ and write $W \models t^-$:*

$$W \models t^+ \quad \text{iff} \quad \forall I \in [\![W]\!] : t \in I$$
$$W \models t^- \quad \text{iff} \quad \forall I \in [\![W]\!] : t \notin I$$

Sightings

| sid | uid | species | date | location | s |
|---|---|---|---|---|---|
| s1 | Carol | bald eagle | 6-14-08 | Lake Forest | − |
| s1 | Carol | fish eagle | 6-14-08 | Lake Forest | − |
| s2 | Alice | raven | 6-14-08 | Lake Placid | + |

Comments

| cid | comment | sid | s |
|---|---|---|---|
| c2 | purple black feathers | s2 | + |

**Figure 3: Belief world "BELIEF Bob" or $\Box_{\text{Bob}}$ of our running example.**

PROPOSITION 7 (POSITIVE AND NEGATIVE BELIEFS). *Let $t$ be a tuple in $Tup_i$, i.e. it is a typed tuple for relation $R_i$. Tuple $t$ is a positive belief for $W$ iff it is in $I^+$. It is a negative belief for $W$ iff it is either in $I^-$ ("stated negative") or if there is another tuple $t' \in I^+$ from $Tup_i$ with the same key ("unstated negative"):*

$W \models t^+$ iff $t \in I^+$
$W \models t^-$ iff $t \in I^- \lor \exists t' \in Tup_i.(t' \in I^+ \land t' \neq t \land key(t') = key(t))$

## 3.2 Belief Databases

A *belief database* is a collection of belief worlds, one for each possible combination of what users believe about the database content or other user's beliefs. We use the notation of multi-modal logic [25] to express *belief statements*. For example, the following statement denotes "Alice believes that Bob believes that tuple $t$ is false":

$$\Box_{\text{Bob}} \Box_{\text{Alice}} t^- \qquad (1)$$

Let $U$ be a set of users. In practice, $U$ is a set of user IDs, but we simply take $U = \{1, \ldots, m\}$. A *belief path* is $w \in U^*$, denoted as $w = w_{[1]} \cdots w_{[d]}$. We further restrict belief paths to be $\in \hat{U}^*$ with $\hat{U}^* = \{w \in U^* \mid w_{[i]} \neq w_{[i+1]}\}$, i.e. belief paths do not contain the same user ids in successive positions. We define a *subpath* as $w_{[i,j]} = w_{[i]} \cdots w_{[j]}$ (defined to be $\varepsilon$ when $i > j$), a *suffix* as a subpath with $w_{[i,d]}$, where $d$ is the *depth* or belief path length of $w$ ($d = |w|$), and we define as usual the concatenation of two sequences $v \cdot w$. We use $\Box_w$, $\Box_{w_{[1,d]}}$, and $\Box_{w_{[1]}} \cdots \Box_{w_{[d]}}$ as equivalent notations. Hence, expression (1) is equal to $\Box_{\text{Bob} \cdot \text{Alice}}\ t^-$.

DEFINITION 8 (BELIEF DATABASE). *(1) A belief statement $\varphi$ is an expression of the form $\Box_w\ t^s$ where $w \in \hat{U}^*$ is a belief path, $t$ is a ground tuple from the tuple universe, and $s \in \{'+', '-'\}$ is a sign.*

*(2) A belief database $D$ is a set of belief statements.*
*(3) Given a belief database $D$ and a belief path $w$. The explicit belief world at $w$ is $D_w = (I^+_w, I^-_w)$ with:*

$$I^+_w = \{t \mid \Box_w\ t^+ \in D\}$$
$$I^-_w = \{t \mid \Box_w\ t^- \in D\}$$

*(4) A belief database $D$ is consistent iff $D_w$ is consistent for all $w \in \hat{U}^*$.*

Figure 2 illustrates the belief database from our running example with eight belief statements. The explicit belief worlds for "Bob believes" and for "Bob believes that Alice believes" are:

$$D_{\text{Bob}} = (\{s2_2, c2_2\}, \{s1_1, s1_2\})$$
$$D_{\text{Bob} \cdot \text{Alice}} = (\{c2_1\}, \emptyset)$$

Continuing this example, lets examine what happens if a new user Dora joins the discussion. Initially there are no belief statements for Dora. In this case, the system needs to assume by default that Dora believes everything that is stated explicitly in the database. If we didn't do so, then we would force Dora to insert explicitly all tuples she agrees with, which are arguably the majority of the tuples in the database. Thus, *by default*, we assume that a user believes every belief statement that is in the database, unless stated otherwise. Dora may later update her belief and disagree explicitly with some tuples; the *default rule* only applies to tuples about which Dora has not expressed explicit disagreement by inserting either a negative belief or a positive belief with the same key but different attributes. For example, when user 1 inserts a belief statement $\Box_1 \, t$, user 2 will believe by default that user 1 believes what he states, i.e. $\Box_{2 \cdot 1} \, t$, but not necessarily the fact itself, i.e. $\Box_2 \, t$. We call this default rule the *message board assumption* in analogy to discussion boards where users state and exchange their opinions about facts and each other beliefs. We define this formally next.

**Definition 9** (Implicit Beliefs). *Given a belief database $D$, define the following sequence $D^{(i)}$:*

$$D^{(0)} = D$$
$$D^{(d+1)} = D^{(d)} \cup \{\Box_i \varphi \mid \varphi \in D^{(d)}, i \in U,$$
$$D^{(d)} \cup \{\Box_i \varphi\} \text{ is consistent}\}$$

**Definition 10** (Theory). *The closure of $D^{(i)}$ is $\bar{D} = \bigcup_{d \geq 0} D^{(d)}$. We call the set $\bar{D}$ the* theory *of $D$.*

The (infinite) belief database $\bar{D}$ captures our intended semantics: it contains all belief statements *explicitly* asserted in $D$ together with all statements that follow *implicitly*, except if they were explicitly contradicted.

**Lemma 11.** *If $D$ is consistent, then $\bar{D}$ is consistent.*

We give now the formal semantics of a belief database, by defining the entailment relationship $D \models \varphi$.

**Definition 12** (Semantics of a Belief Database). *A belief database $D$ entails a belief statement $\varphi$, in notation $D \models \varphi$, if $\varphi \in \bar{D}$.*

We illustrate with our running example (Fig. 2). After Carol inserted her statement $(i_1\text{:}s_1^+)$, Alice and Bob believe the bald eagle sighting by default ($D \models \Box_{\text{Alice}} \, s_1^+$). Bob, however, does not want to believe this sighting and explicitly states his disagreement ($i_2$: $\Box_{\text{Bob}} \, s_1^-$). While he does not believe it himself, he still believes that Alice believes this sighting ($D \models \Box_{\text{Bob} \cdot \text{Alice}} \, s_1^+$).

### 3.3 Queries over Belief Databases

We now introduce our language for querying belief databases which consists of conjunctive queries extended with belief annotations. We call these *Belief Conjunctive Queries* (BCQ) and adopt a compact, Datalog-like syntax that combines elements from multi-modal logic.

**Definition 13** (BCQ syntax). *A belief conjunctive query is an expression of the form*

$$q(\bar{x}) :\text{--} \Box_{\bar{w}_1} R_1^{s_1}(\bar{x}_1), \ldots, \Box_{\bar{w}_g} R_g^{s_g}(\bar{x}_g) \ ,$$

*consisting of a query head $q(\bar{x})$ and $g$ belief atoms or* modal subgoals *forming the query body. Each modal subgoal $\Box_{\bar{w}_i} R_i^{s_i}(\bar{x}_i)$ comprises a belief path $\bar{w}_i$, a sign $s_i$, and a relational atom $R_i(\bar{x}_i)$ with relational tuples $\bar{x}_i$.*

We call a modal subgoal $\Box_{\bar{w}} R^s(\bar{x})$ positive if $s = '+'$, and negative if $s = '-'$. We write $\bar{x}$ and $\bar{w}$ for tuples and belief paths. They can contain both variables and constants. We write $var(\bar{w})$ and $var(\bar{x})$ to denote the variables of $\bar{w}$ and $\bar{x}$. We also allow *arithmetic predicates* in the query body, using standard operators $\neq, <, >, \leq$, and $\geq$. A variable occurrence in a belief path or a positive relational atom is called a *positive occurrence*. A query is *safe* if every variable has at least one positive occurrence. We assume all queries to be safe.

We define next the semantics of a query. We write below $D \models \varphi_1, \ldots, \varphi_g$ for $\bigwedge_i (D \models \varphi_i)$, where $\varphi_1, \ldots, \varphi_g$ are belief statements.

**Definition 14** (BCQ semantics). *Let $q$ be a query with head variables $\bar{x}$ and body variables $\Phi$. The answer to $q$ on a belief database $D$ is the following set of tuples over the set of constants in the attribute domains:*

$$\{\theta(\bar{x}) \mid \theta : var(\Phi) \mapsto const, D \models \theta(\Phi)\}$$

In other words, for every valuation $\theta$ that maps variables to constants in the attribute domains, consider the formula $\theta(\Phi)$, which is of the form $\varphi_1, \ldots, \varphi_g$ (one belief statement for each subgoal): if $D$ entails $\theta(\Phi)$, then we return the tuple $\theta(\bar{x})$. Recall that a belief world can entail positive and negative beliefs (Def. 6). Depending on its sign $s$ and its belief path $\bar{w}$, each subgoal represents positive or negative beliefs of one or more belief worlds. A BCQ then asks for constants in relational tuples and belief paths that imply positive beliefs in positive subgoals, and negative beliefs in negative subgoals.

**Example 15.** *Using $S$ for the relation Sightings, the following query returns all users $x$ who disagree with any of Alice's beliefs, i.e. who have a negative belief about some tuple $(y, z, u, v, w)$, which is a positive belief for Alice at the same time.*

$$q_3(x)\text{:--}\Box_x S^-(y,z,u,v,w), \Box_{\text{'Alice'}} S^+(y,z,u,v,w)$$

### 3.4 Discussion

Default rules like our message board assumption are studied in *default logics*. In our presentation, we avoided introducing default logics, non-monotonic reasoning, and stable model semantics, and opted for a simpler definition. Yet, an alternative formulation of our message board assumption can be given using Reiter's default logic [41]: The set of formulas $\bar{D}$ that we define in Def. 9 is provably equal to the provably unique stable model for $D$ under the default rule (see appendix C):

$$d_s = \frac{\varphi : \Box_i \varphi}{\Box_i \varphi}$$

Designing an appropriate data and query model for belief databases requires a fine tradeoff between tractability and expressiveness. Reasoning in modal logics can quickly become intractable [26]. This applies, in particular, to fragments that include *possibility* in addition to *certainty* and *impossibility* (positive or negative beliefs). In the notation of modal logics, we allow statements of the form $\Box_{\text{Alice}} t$ and $\Box_{\text{Alice}} \neg t$ (Alice believes that $t$ is necessary or impossible). Complexity would considerable increase by allowing negations before the modal operators, e.g. $\neg \Box_{\text{Alice}} t$ (Alice does not believe that $t$ is necessary), which is equal to $\Diamond_{\text{Alice}} \neg t$ (Alice believes that $\neg t$ is possible). In our fragment of modal logics, we allow negations only on ground facts, noting that this is sufficient to express conflicts.

The general approach for defining semantics in modal logic is through axioms and *Kripke structures* [21, 25]. Every concrete logic consists of a class of axioms and considers formulas that are logical consequences from these axioms, where entailment is defined in terms of *Kripke structures*. Often, axioms can be removed by restricting the class of Kripke structures. For example, the axioms in K5 are equivalent to restricting Kripke structures to have accessibility relations that are symmetric and transitive. We have chosen to define the semantics of a belief database without the aid of axioms and Kripke structures, because we felt it is simpler for our setting. On the other hand, our definition does not lead to an obvious query evaluation procedure. To derive such a procedure we introduce a particular Kripke structure next, and show that it defines a semantics that is equivalent to that in Def. 12.

## 4. CANONICAL KRIPKE STRUCTURE

We review here Kripke structures [25], then define our canonical Kripke structure that captures precisely the semantics of belief databases (Def. 12).

A *rooted Kripke structure* is $K = (V, (W_v)_{v \in V}, (E_i)_{i \in U}, v_0)$ where:
- $V$ is a finite set called *states*,
- $W_v = (I_v^+, I_v^-)$ is a *belief world* associated with each state $v \in V$,
- $E_i \subseteq V \times V$ is a set of *edges* or accessibility relations associated with each user $i \in U$,
- $v_0 \in V$ is the *root* of the Kripke structure.

Given a rooted Kripke structure $K$ and a state $v$, the entailment relationship $(K, v) \models \varphi$ is defined recursively as:

$$(K, v) \models t^+ \quad \text{if } W_v \models t^+ \quad \text{(Def. 6)}$$
$$(K, v) \models t^- \quad \text{if } W_v \models t^- \quad \text{(Def. 6)}$$
$$(K, v) \models \Box_i \varphi \quad \text{if } \forall (v, v') \in E_i. (K, v') \models \varphi$$

We write $K \models \varphi$ if $(K, v_0) \models \varphi$.

We illustrate with the Kripke structure of Fig. 4. There are four states $\#0, \ldots, \#3$ with the root $\#0$. Consider the belief world at state $\#2$, $W_{\#2} = (I_{\#2}^+, I_{\#2}^-)$. $I_{\#2}^+$ consists of the tuples $s2_2, c2_2$ and $I_{\#2}^-$ of the tuples $s1_1, s1_2$. We therefore have $(K, \#2) \models s2_2$. As all edges labeled 2 from the root lead to the state $\#2$, we further have $K \models \Box_2 s2_2$. In the following, we use interchange-ably the notions of *world id* (e.g. $\#3$) or *belief path* (e.g. $w = 2 \cdot 1$), and those of *state* or *world*.

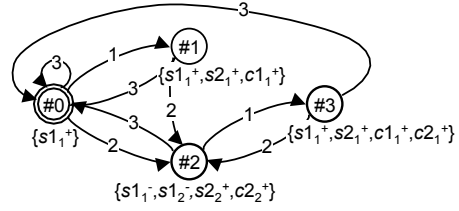

**Figure 4: The canonical Kripke structure for our running example.**

Consider a belief database $D$. We define the *support states* as the set of all belief paths $w$ for which $D$ contains a belief statement over $w$, and the *states* as the set of all their prefixes:

$$Supp(D) = \{w \in \hat{U}^* \mid D_w \neq (\emptyset, \emptyset)\}$$
$$States(D) = \{w \in \hat{U}^* \mid \exists u \in U^* : w \cdot u \in Supp(D)\}$$

For any $w \in \hat{U}^*$, we define the *suffix states* as all the suffixes of $w$ that are in $States(D)$, and the *deepest suffix state* ($dss$) as the suffix state with the longest belief path:

$$SuffixStates(w) = \{v \in States(D) \mid \exists u \in U^* : u \cdot v = w\}$$
$$dss(w) = \text{max-arg}_v \{|v| \mid v \in SuffixStates(w)\}$$

We can now define formally the canonical Kripke structure for a belief database $D$:

DEFINITION 16 (CANONICAL KRIPKE STRUCTURE). *Let $D$ be a belief database, and denote $V = States(D)$. The canonical Kripke structure is $K(D) = (V, (\bar{D}_v)_{v \in V}, (E_i)_{i \in U}, \varepsilon)$, with edges defined as:*

$$E_i = \{(w, dss(w \cdot i)) \mid w \in States(D), w \cdot i \in \hat{U}^*\}$$

We describe informally the canonical Kripke structure for $D$. Start with all the belief paths $w$ that are mentioned in some belief statement in $D$: these form the support states. Take all their prefixes: these form all states of $K(D)$. Next, for each state $v$, compute the belief world $\bar{D}_v$: this is the belief world for $v$ in the closure of $D$. Although the closure $\bar{D}$ is an infinite object, the set $\bar{D}_v$ is contained in $D^{(d)}$ where $d = |v|$. Thus, in order to compute $\bar{D}_v$ it suffices to compute $D^{(d)}$ through a finite process, then take $\bar{D}_v = D_v^{(d)}$. Finally, edges labelled $i$ in $K(D)$ go "forward" from a state $w$ to state $w \cdot i$ if the latter exists. Otherwise they go "back" to the state with the longest belief path that is a suffix of the desired, but missing state $w \cdot i$. That means, edges labelled $i$ always go from a state $w$ to $dss(w \cdot i)$.

We prove the following theorem in the appendix:

THEOREM 17 (CANONICAL KRIPKE STRUCTURE). *(1) For any belief statement $\varphi$, $D \models \varphi$ iff $K(D) \models \varphi$. (2) $K(D)$ can be computed in time $O(m^d n)$, where $n$ is the size of the belief database $D$, $m$ the number of users and $d$ is the maximum depth of any belief path in $D$.*

## Figure 5

| Sightings_STAR | | | | | |
|---|---|---|---|---|---|
| *tid* | *sid* | *uid* | *species* | *date* | *location* |
| s1.1 | s1 | 3 | bald eagle | 6-14-08 | Lake Forest |
| s1.2 | s1 | 3 | white eagle | 6-14-08 | Lake Forest |
| s2.1 | s2 | 1 | crow | 6-14-08 | Lake Placid |
| s2.2 | s2 | 1 | raven | 6-14-08 | Lake Placid |

| Comments_STAR | | | |
|---|---|---|---|
| *tid* | *cid* | *comment* | *sid* |
| c1.1 | c1 | found feathers | s2 |
| c2.1 | c2 | black feathers | s2 |
| c2.2 | c2 | purple black feathers | s2 |

| Comments_V | | | | |
|---|---|---|---|---|
| *wid* | *tid* | *cid* | *s* | *e* |
| 1 | c1.1 | c1 | + | y |
| 2 | c2.2 | c2 | + | y |
| 3 | c1.1 | c1 | + | n |
| 3 | c2.1 | c2 | + | y |

| Users | |
|---|---|
| *uid* | *name* |
| 1 | Alice |
| 2 | Bob |
| 3 | Carol |

| Sightings_V | | | | |
|---|---|---|---|---|
| *wid* | *tid* | *sid* | *s* | *e* |
| 0 | s1.1 | s1 | + | y |
| 1 | s1.1 | s1 | + | n |
| 1 | s2.1 | s2 | + | y |
| 2 | s1.1 | s1 | − | y |
| 2 | s1.2 | s1 | − | y |
| 2 | s2.2 | s2 | + | y |
| 3 | s1.1 | s1 | + | n |
| 3 | s2.1 | s2 | + | n |

| E | | |
|---|---|---|
| *wid1* | *uid* | *wid2* |
| 0 | 1 | 1 |
| 0 | 2 | 2 |
| 0 | 3 | 0 |
| 1 | 2 | 2 |
| 1 | 3 | 0 |
| 2 | 1 | 3 |
| 2 | 3 | 0 |
| 3 | 2 | 2 |
| 3 | 3 | 0 |

| D | |
|---|---|
| *wid* | *d* |
| 0 | 0 |
| 1 | 1 |
| 2 | 1 |
| 3 | 2 |

| S | |
|---|---|
| *wid1* | *wid2* |
| 1 | 0 |
| 2 | 0 |
| 3 | 1 |

**Figure 5: Relational representation of the canonical Kripke structure for our running example.**

Note that $K(D)$ encodes an infinite number of belief worlds with a finite number of states. This provides the basis for our query evaluation approach: given a belief database $D$, compute its canonical Kripke structure $K(D)$, then evaluate queries over $K(D)$. We address the latter step in the next section.

## 5. TRANSLATION

This section covers the representation of belief databases in the standard relational model. In particular, we give (1) the representation of the canonical Kripke structure, (2) the translation of belief conjunctive queries over this representation, and (3) updates of a database.

### 5.1 The relational representation

The relational representation uses an *internal schema* $\mathcal{R}^* = (R_1^*, \ldots R_r^*, U, V_1, \ldots V_r, E, D, S)$. Recall that the first attribute of each content relation $R_i(\overline{att_i})$ contains the *external key* attribute $key_i$ of that relation. Each relation $R_i$ is represented by an internal relation $R_i^*$ with one additional attribute $tid$ and the relation obeying the functional dependency $tid \rightarrow Attr(R_i)$. The *internal key* constraint is only on this surrogate key: $R_i^*(\underline{tid}, key_i, att_{i2}, \ldots, att_{l_i})$. In addition, the internal schema includes: a user relation $U$ with user ids and optional user attributes; $r$ valuation relations $V_i$, one for each $R_i$, recording tuples, their keys, signs and whether they are explicit or implicit in each belief world (explicit means explicitly annotated in contrast to implicitly inferred by the default assumption); an edge relation $E$ containing the accessibility relations between worlds for each user; a depth relation $D$ recording the nesting depth of each world id; and a suffix relation $S$ recording the deepest suffix state for each world. Relations $D$ and $S$ record information that is used during updates of the database.

The representation of the canonical Kripke structure is then straight forward: For each world $w \in States(D)$ we create a unique world identifier $wid(w)$ and insert it into relation $D$ together with its nesting depth:

$$D(wid(w), |w|)$$

Analogously, create one entry in relation $S$ that records the deepest suffix state for each world:

$$S(wid(w), wid(dss(w_{[2,d]})))$$

Each tuple $R_i(k, x_2, \ldots, x_{l_i})$ of any world is inserted as

$$R_i^*(t, k, x_2, \ldots, x_{l_i}) ,$$

where $t$ is its unique internal key. Note that $R_i^*$ gathers tuples from $R_i$ of all worlds. All worlds in which $t$ appears are recorded in the valuation relation $V_i$ as

$$V_i(wid(w), t, k, s, e) ,$$

where $s$ is its sign '+' if positive or '−' if negative, and $e$ is 'y' or 'n' depending on whether the tuple is explicit or not in the particular world. This attribute indirectly records the "provenance" for each tuple in a world (explicitly asserted or implicitly inferred by the message board assumption) and is needed during updates; it implicitly tracks the origin world of an implicit tuple and allows to determine precedence in case of updates with inconsistent values. The external key $k$ is included in the valuation relations in order to detect conflicts between different belief worlds by merely inspecting the valuation relations and, thereby, to increase efficiency during updates. Finally, for each $(u, v) \in E_j$, insert an entry into relation $E$:

$$E(u, j, v)$$

Figure 5 shows the representation of our running example. Attributes *wid* stand for world id, *tid* for tuple id, *uid* for user id, *s* for sign, *e* for explicitness, and *d* for nesting depth.

### 5.2 Query translation

We next describe the translation of any belief conjunctive query into non-recursive Datalog over the internal schema: The translation first creates one temporary tables for each subgoal and then creates one query over these tables (Algorithm 1).

Recall that a BCQ consists of $g$ positive or negative modal subgoals and optional additional arithmetic atoms (Def. 13). Conceptually, each positive subgoal represents a subquery for positive beliefs, and each negative subgoal for negative beliefs. A belief conjunctive query then asks for constants in relational tuples and belief paths that imply positive beliefs in positive subgoals, and negative beliefs in negative subgoals. Also recall from Prop. 7 that a negative belief can be either *stated negative*, i.e. due to an explicitly stated negative belief $t^-$, or *unstated negative*, i.e. due to an explicitly stated positive belief $t'^+$, where tuple $t'$ has the same key as $t$. Both of these cases have to be considered during query translation, which makes the translation for negative subgoals more complex, requiring nested disjunctions with negation. Also note that a negative

**Algorithm 1**: Translation of any BCQ over the canonical belief representation.

**Input**: BCQ $q(\bar{x}) :- \Box_{\bar{w}_1} R_1^{s_1}(\bar{x}_1), \ldots, \Box_{\bar{w}_g} R_g^{s_g}(\bar{x}_g)$
**Output**: Translated query $Q(\bar{x})$ over temporary tables

1. Check safety of query:
$\forall \alpha \in var(q) : \alpha \in \left(\bigcup_i var(\bar{w}_i)\right) \cup \left(\bigcup_{i.(s_i='+')} var(\bar{x}_i)\right)$
2. For each subgoal $i$, create a temporary table $T_i$:
$T_i(\bar{w}_i, \bar{x}, s) :- E^*(0, \bar{w}_i, z), V(z, t, \_, s, \_), R_i^*(t, \bar{x})$
3. Compose the final query with one temporary table $T_i$ and one condition $C_i$ for each subgoal $i$ ...
$q(\bar{x}) :- T_1(\bar{w}_{t1}, \bar{x}_{t1}, s_{t1}), \ldots, T_g(\bar{w}_{tg}, \bar{x}_{tg}, s_{tg}), C_1, \ldots, C_g$
4. ... where conditions for positive subgoals are:
$C_i = (\bigwedge_{j:1-d_i} \bar{w}_{ti[j]} = \bar{w}_{i[j]}), s_{ti} = 1,$
$\bigwedge_{j:1-l_i} \bar{x}_{ti[j]} = \bar{x}_{i[j]},$
5. ... and conditions for negative subgoals are:
$C_i = (\bigwedge_{j:1-d_i} \bar{w}_{ti[j]} = \bar{w}_{i[j]}), \bar{x}_{ti[1]} = \bar{x}_{i[1]},$
$\left((s_{ti} = 0, \bigwedge_{j:2-l_i} \bar{x}_{ti[2,j]} = \bar{x}_{i[2,j]}) \right.$
$\left. \vee (s_{ti} = 1, \bigvee_{j:2-l_i} \bar{x}_{ti[2,j]} \neq \bar{x}_{i[2,j]})\right),$

subgoal alone is unsafe, since a single positive tuple in a belief world implies negative beliefs for all tuples from the same tuple universe with the same key.

The algorithm first verifies safety: each variable of the query has to appear in a belief path or the relational tuples of a positive subgoal (1). It then creates, for each subgoal $\Box_{\bar{w}_i} R_i^{s_i}(\bar{x}_i)$, a temporary table $T_i$ (2), with $E^*(y, \bar{w}, z)$ being a notational shortcut for

$$E^*(y, \bar{w}, z) \stackrel{\text{def}}{=} E(y, \bar{w}_{[1]}, z_1), \ldots, E(z_{d-1}, \bar{w}_{[d]}, z),$$

with $z \stackrel{\text{def}}{=} y$ for $w = \varepsilon$ This table has arity $l_i + d_i + 1$ and includes all stated tuples for all worlds with belief path $\bar{w}$. Recall that $\bar{w}$ can have both constants and variable, so that an intermediate table can encode the valuations for more than one belief world. Note that we cannot perform arbitrary selections and projections for negative subgoals at this point, even if $\bar{x}$ includes constants. Any positive tuple can lead to another tuple being impossible that may actually be required to be joined with another positive or negative subgoal.

The final query (3) then combines those tables as follows: For positive subgoals, it choses positive stated tuples ($s = 1$), and choses constants or joins to other subgoals (4). For negative subgoals (5), it distinguishes the case of stated impossible tuples, i.e. $s = 0$, and unstated impossible tuples, i.e. positive tuples with $s = 1$ that share the key to at least another certain tuple in another positive subgoal. Arithmetic predicates are simply added as additional condition to the translated query.

The following example illustrates this translation.

EXAMPLE 18. *Assume a relation $R(\underline{sample}, category, origin)$ that classifies empirical samples into a number of categories and records their origin. Consider a query for disputed samples, i.e. samples $x$ for which at least two users $y$ and $z$ disagree on its category or origin:*

$$q(x, y, z) :- \Box_y R^+(x, u, v), \Box_z R^-(x, u, v)$$

*The query written in BeliefSQL is:*

```
select    R1.sample, U1.name, U2.name
from      Users as U1, Users as U2
          BELIEF U1.uid R as R1,
          BELIEF U2.uid not R as R2,
where     R1.sample = R2.sample
and       R1.category = R2.category
and       R1.origin = R2.origin
```

*The translation over the canonical belief representation first creates two intermediate tables:*

$T_1(y, x, u, v, s) :- E(0, y, z_1), V(z_1, t, x, s, \_), R^*(t, \_, u, v)$
$T_2(z, x, u, v, s) :- E(0, z, z_1), V(z_1, t, x, s, \_), R^*(t, \_, u, v)$

*The final query then combines those two tables*

$Q(x, y, z) :- T_1(y, x, u, v, '+'), T_2(z, x, u_2, v_2, s_2),$
$(s_2 = '-' \wedge u_2 = u \wedge v_2 = v) \vee (s_2 = '+' \wedge (u_2 \neq u \vee v_2 \neq v))$

## 5.3 Updates

Updates on a belief database consist of several smaller, often conditional operations; those operations often incorporate the result of non-recursive queries extended with a `max`-operator over the existing data. As a compact notation for these updates, we write in the following $\triangle R$ and $\triangledown R$ to refer to a set of tuples that are inserted into or deleted from $R$:

$$R^{\text{new}} = (R^{\text{old}} - \triangledown R) \cup \triangle R$$

We again use the letter $T$ for temporary tables, and use expressions of the form $\exists (\_, \_, z) \in T$ as notational shortcut for $\exists x, y.(x, y, z) \in T$. In order to specifically refer to keys, we write $R(k, \bar{x}')$ for relational tuples, where $\bar{x}'$ refers to $\bar{x}_{[2,l]}$ in $R(\underline{k}, x_2, \ldots, x_l)$.

**Data inserts.** Assume a desired insert $\Box_w R^s(k, \bar{x}')$, i.e. we want to insert a tuple $R(k, \bar{x}')$ with sign $s$ into world $w$. Such an insert first has to assure that the world $w$ already exists before the tuple can be inserted. Algorithm 2 (idWorld) does so by verifying that the path $w$ from the root leads to a world at depth $d = |w|$ (1). If not, it recursively verifies that its parent node exists (3). Note that complexitywise, this recursion can be unfolded as it happens a maximum of $d$ times. idWorld then creates a new world id and applies necessary operations on the canonical model (4- 7). One such operation finds the deepest suffix state (dss) of a world (Algorithm 3). This procedure needs the `max`-operator. The back link to the $dss(w)$ is stored in relation $S$ (8). After creating a new world, idWorld inserts all tuples from $dss(w)$ as implicit tuples (9).

Given the world id $y$ of $w$, insertTuple (Algorithm 4) first verifies if the tuple $(\_, k, \bar{x}')$ already exists in $R^*$; if not, it creates a new entry (1). It then inserts the tuple into world $w$ only if this update is consistent with existing explicit beliefs (5). If inserted, insertTuple also has to verify possible updates in all *dependent worlds* of $w$ (8). Dependent worlds are those for which $x$ is a suffix state. In order of increasing depth, it verifies for each dependent world $z$ (9) that an update has no explicit conflict in $z$ (12) and no conflict in the $dss$ of $z$ (14). If there are no such conflicts, the tuple gets inserted and overwrites any existing implicit conflicting beliefs.

**Algorithm 2**: (idWorld) Returns the identifier $x$ of a world $w$. Creates new world if it does not exist yet.

**Input**: World belief path $w$
**Output**: World id $x = wid(w)$

1. Define $d = |w|$; check that depth of $x$ is $d$:
   $T(x) \colon\!\!- E^*(0, w, x), D(x, d)$
2. **if** $T$ *is empty* **then**
3.     Get the parent id:
   $x' = \mathsf{idWorld}(w_{[1,d-1]})$
4.     Create a new id $x$ for $w$ and a new entry in $D$:
   $\triangle D(x, d)$
5.     Redirect the $w_{[d]}$-edge from $x'$ to $x$:
   $\triangledown E(x', w_{[d]}, \_), \triangle E(x', w_{[d]}, x)$
6.     For all users $u$ except $w_{[d]}$, create a $u$-edge from $x$ to the deepest suffix state of $w \cdot u$:
   $\triangle E(x, u, \mathsf{dss}(w \cdot u)) \colon\!\!- U(u, \ldots), u \ne w_{[d]}$
7.     For all worlds $v \cdot w_{[1,d-1]}$ for which $w$ is the deepest suffix state for $v \cdot w$, update the $w_{[d]}$-edge:
   $\triangledown E(y, w_{[d]}, \_) \colon\!\!- E^*(v, w_{[1,d-1]}, y), D(y, r), r \geq d,$
   $\qquad E(y, w_{[d]}, z), D(z, p), p < d$
   $\triangle E(y, w_{[d]}, x) \colon\!\!- E^*(v, w_{[1,d-1]}, y), D(y, r), r \geq d,$
   $\qquad E(y, w_{[d]}, z), D(z, p), p < d$
8.     Create backlink to deepest suffix in $S$:
   $\triangle S(x, \mathsf{dss}(w_{[2,d]}))$
9.     Insert all implicit tuples into new world $w$:
   $\triangle V(x, t, y, s, \text{'n'}) \colon\!\!- S(x, z), V(z, t, y, s, \_)$
10. **return** $x$

---

**Algorithm 3**: (dss) Returns the world id of the deepest suffix state for belief path $w$.

**Input**: World belief path $w$
**Output**: $z = wid(dss(w))$

1. Query ids $z$ and depths $d$ of all suffix worlds:
   **for** $p = 1 \ldots (d+1)$ **do**
   $\quad T(z, y) \colon\!\!- E^*(0, w_{[p,d]}, z), D(z, y)$
2. Return the id $z$ of the world with maximum depth:
   **return** $z$ from $T(z, d)$ where $d = \mathtt{max}(d)$

---

**Other updates.** For a new user insert, first a new entry in relation $U$ with a unique $uid$ has to be added: $\triangle U(u, \ldots)$. Then, back edges from each world to the root have to be added: $\triangle E(x, u, 0) \colon\!\!- D(x, \_)$. Delete operations follow a similar semantics as inserts.

## 5.4   Space complexity

We next give theoretic bounds for the size of a BDMS in the number of tuples in the underlying RDBMS. Let $m$ be the number of users, $n$ be the number of annotations, $\bar{d}$ the average depth of belief annotations, and $N$ the number of states in the canonical Kripke structure. Sizes of relations are $|U| = m$, $|D| = N$, $|S| = N - 1$, $|R^*| = \mathcal{O}(n)$, and $|E| = \mathcal{O}(mN)$. An insert into world $w$ can create up to $N_w^S$ entries in table $V$, where $N_w^S$ is 1 plus the number of worlds for which $w$ is a suffix state. For the root $\varepsilon$, $N_\varepsilon^S = N$, and hence, an insert at the root can create up to $N$ inserts into $V$. Hence, $|V| = \mathcal{O}(nN)$, and the overall database size $|\mathcal{R}^*| = \mathcal{O}((n+m)N)$.

In theory, $N$ is only loosely bounded by $\mathcal{O}(n\bar{d})$ with the average depth of annotations $\bar{d}$ as the number of

---

**Algorithm 4**: (insertTuple) Inserts signed tuple $R^s(k, \bar{x}')$ into existing world $w$ if insert is consistent. Returns the success of insert attempt.

**Input**: World belief path $w$ and id $y$, signed tuple $R^s(k, \bar{x}')$
**Output**: *Success*

1. Get existing or create new internal key $t$ for tuple $R(k, \bar{x}')$:
   $\triangle R^*(t, k, \bar{x}')$
2. Get all tuples of world $y$ with key $k$:
   $T_1(t', s', e') \colon\!\!- V(y, t', k, s', e')$
3. If $t^s$ is already explicitly present in the world:
   **if** $(t, s, \text{'y'}) \in T_1$ **then return** *false*
4. If $t^s$ is already implicitly present in the world:
   **if** $(t, s, \text{'n'}) \in T_1$ **then**
   $\quad \triangledown V(y, t, k, s, \text{'n'}), \triangle V(y, t, k, s, \text{'y'})$, **return** *true*
5. If $t$ does not conflict with an existing explicit tuple ...
   **if** $s = \text{'+'} \land \not\exists (t, \text{'-'}, \text{'y'}) \in T_1 \land \not\exists (\_, \text{'+'}, \text{'y'}) \in T_1$ **or**
   $s = \text{'-'} \land \not\exists (t, \text{'+'}, \text{'y'}) \in T_1$ **then**
6.     ... delete any conflicting implicit tuples:
   **if** $s = \text{'+'}$ **then** $\triangledown V(y, t, k, \text{'-'}, \text{'n'}), \triangledown V(y, \_, k, \text{'+'}, \text{'n'})$
   **if** $s = \text{'-'}$ **then** $\triangledown V(y, t, k, \text{'+'}, \text{'n'})$
7.     ... insert $t^s$ into $y$:
   $\triangle V(y, t, k, s, \text{'y'})$
8.     ... get all dependent worlds of $w$ and their depth:
   $T_2(z, r) \colon\!\!- E^*(\_, w, z), D(z, r), r > d$
9.     ... then, for each dependent world $z$ in order of depth:
   **foreach** $z \in T_2$ *in ascending order of* $r$ **do**
10.       Get all tuples of world $z$ with key $k$:
    $T_3(t'', s'', e'') \colon\!\!- V(z, t'', k, s'', e'')$
11.       Insert $t^s$ into world $z$ if there is no conflict:
    **if** $s = \text{'+'} \land \not\exists (t, \text{'-'}, \_) \in T_3 \land \not\exists (\_, \text{'+'}, \_) \in T_3$ **or**
    $s = \text{'-'} \land \not\exists (t, \text{'+'}, \_) \in T_3$ **then** $\triangle V(z, t, x_1, s, \text{'n'})$
12.       Otherwise if conflicts are not explicit:
    **else if** $s = \text{'+'} \land \not\exists (t, \text{'-'}, \text{'y'}) \in T_3 \land \not\exists (\_, \text{'+'}, \text{'y'}) \in T_3$
    **or** $s = \text{'-'} \land \not\exists (t, \text{'+'}, \text{'y'}) \in T_3$ **then**
13.         Get tuples with key $k$ from $dss(z)$:
    $T_4(t''', s''') \colon\!\!- S(z, v), V(v, t''', k, s''', \_)$
14.         Update $z$ if there are no conflicts with $dss(z)$:
    **If** $s = \text{'+'} \land \not\exists (t, \text{'-'}) \in T_4 \land \not\exists (\_, \text{'+'}) \in T_4$ **then**
    $\triangledown V(y, t, k, \text{'-'}, \text{'n'}), \triangledown V(y, \_, k, \text{'+'}, \text{'n'})$
    $\triangle V(y, t, k, \text{'+'}, \text{'n'})$
    **If** $s = \text{'-'} \land \not\exists (t, \text{'+'}) \in T_4$ **then**
    $\triangledown V(y, t, k, \text{'+'}, \text{'n'})$,
    $\triangle V(y, t, k, \text{'-'}, \text{'n'})$
    **return** *true*
    **else**
    $\quad$ **return** *false*

---

prefixes, hence, possible states. However, for bounded nesting depth of belief paths ($|w| \leq d_{\max}$), we have $N = \mathcal{O}(m^{d_{\max}})$ as the number of possible different belief paths $|\hat{U}^{d_{\max}}|$ with depth up to $d_{\max}$, which is constant in $n$. We then have $|\mathcal{R}^*| = \mathcal{O}((n+m)m^{d_{\max}})$, which becomes $\mathcal{O}(n \cdot m^{d_{\max}})$ for $n >> m$.

We call the factor $\frac{|\mathcal{R}^*|}{n}$ the *relative overhead* in size for adding beliefs to databases. We have seen above that this factor is $\mathcal{O}(m^{d_{\max}})$ in the worst case, which is quite significant. For example, it is around 10,000 for a belief database with $m = 100$ users and belief annotations of depth up to $d_{\max} = 2$. In practice, however, the overhead heavily depends on the number of belief worlds affected by inserts, which, in turn, depends on skews in the underlying annotations. We will illustrate these effects on the size of a BDMS by varying parameters in the annotation data in the next section.

Table 1: Relative overhead $\frac{|\mathcal{R}^*|}{n}$ of the size of a belief database for $n = 10,000$ annotations, 10 or 100 users, varying user participation (Zipf or uniform) and 3 distributions of annotation depth.

|  | $m = 10$ | | $m = 100$ | |
|---|---|---|---|---|
| $\Pr[d = \{0,1,2\}]$ | Zipf | uniform | Zipf | uniform |
| $[0.\bar{3}, 0.\bar{3}, 0.\bar{3}]$ | 31 | 38 | 130 | 1,009 |
| $[0.8, 0.19, 0.01]$ | 27 | 60 | 68 | 162 |
| $[0.199, 0.8, 0.001]$ | 7 | 6 | 21 | 26 |

## 6. EVALUATION

We have implemented a prototype of a BDMS that allows bulk insertions of belief annotations and translations of belief conjunctive queries into SQL. We use this prototype to experimentally study (i) the relative overhead of managing annotations and (ii) query performance. The program to generate annotations and to translate queries is implemented in Java and uses JDBC for calls to a RDBMS. As experimental platform, we run Microsoft SQL server 2005 on a Dual-Xeon machine (3GHz) with 4G of main memory. We use the database schema from our running example of Fig. 5, neglecting the comments table for the experiments: $\mathcal{R}^* = (S^*, U, V_S, E, L, H)$ and measure the size as the number of all tuples in the database ($|\mathcal{R}^*|$). Clustered indexes are available over the internal keys. All experiments are performed on synthetic data.

### 6.1 Size of a BDMS

We have seen in Sect. 5.4 that the *relative overhead* for adding beliefs to a database ($\frac{|\mathcal{R}^*|}{n}$), i.e. the number of tuples in the database per number of belief annotations, is $\mathcal{O}(m^{d_{\max}})$, which is 100 for $m = 10$ and $10,000$ for $m = 100$ users, and belief annotations of maximum depth $d_{\max} = 2$. In practice, the skew in the annotation, i.e. the distribution of the path length $k$, and the distinct count of belief paths can reduce this overhead dramatically. To study this dependency, we use a generic annotation generator that creates parameterized belief annotations. We model annotation skew as discrete probability distributions $\Pr[k = x]$ of the nesting depth of annotations (e.g. 1% of annotations are of nesting depth 2) and user participation as either uniform or following a generalized Zipf distribution (e.g. user 1 is responsible for 50% of all annotations, user 2 for 25%, ...). Table 1 shows the relative overhead of synthetic belief databases (each value averaged over 10 databases with the same parameters) and illustrates its variations with different distributions. Figure 6 further shows that the relative overhead can actually increase or decrease with the number of annotations $n$. The decrease for the lower more skewed distribution arises from the decreasing relative overhead for supporting a constant number of users $m$ for increasing $n$: $\mathcal{O}(\frac{n+m}{n} m^{d_{\max}})$. Also note that, despite the upper blue graph suggesting an exponentially increasing relative overhead, it flattens again and will not surpass its theoretic bound of 10,000 in the limit. The take-away of this experiment is that the actual overhead of belief annotations can be significantly lower than their theoretic bound. But it is still substan-tial and efficient techniques are needed to create more compact representations of belief databases. We shortly discuss future work on alternative representations at the end of this section.

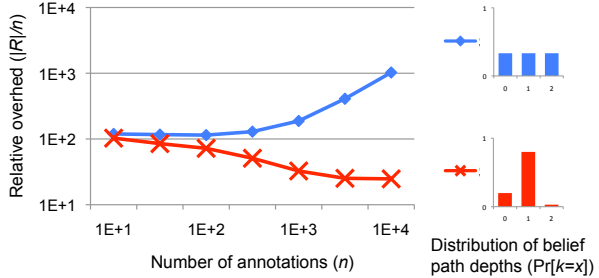

Figure 6: $\frac{|\mathcal{R}^*|}{n}$ crucially depends on the annotation skew and can either increase or decrease with $n$ (100 users with uniform participation).

### 6.2 Query complexity

In the following, we list 3 example queries. These queries cover the typical usage patterns in a BDMS and illustrate the enriched query semantics it can support.

1. The first type is a *query for content*. It asks for the content of a particular belief world and is of the form "What does Alice believe?" In addition, we vary the depth of its belief path $d \in \{0, \ldots, 4\}$:

   $q_{1,d}(x,y) :- \Box_w S^+(x,\_,y,\_,\_)$, with $|w| \in \{0,\ldots,4\}$

2. The second type is a *query for conflicts*. It asks for conflicts between belief worlds and corresponds to: "Which animal sightings does Bob believe that Alice believes, which he does not believe himself?"

   $q_2(x,y) :- \Box_{2 \cdot 1} S^+(x,z,y,u,v), \Box_2 S^-(x,z,y,u,v)$

3. The third query is an example of a *query for users*, i.e. a query that explicitly includes a user id as variable in the answer. It corresponds to: "Who disagrees with any of Alice's beliefs of sightings at Lake Placid?" Note that the query variable only appears in the belief path of a negative subgoal.

   $q_3(x) :- \Box_x S^-(y,z,u,v,\text{'a'}), \Box_1 S^+(y,z,u,v,\text{'a'})$

The evaluation time scales roughly linear with the size of the BDMS ($|\mathcal{R}^*|$). In Table 2 we report the size of the result sets, average query times and standard deviation for a belief database with 10,000 annotations and 224,339 tuples (relative overhead 22.4). Each query was executed 1,000 times. Before each query execution, we clear all database caches with SQL server specific commands. Remaining variations in execution times of identical queries result from fluctuations in the OS beyond our control. The runtimes are in the hundreds of milliseconds. Content query $q_1$ is clearly fastest as it ranges only over one world. The execution time increases by adding 1 join with relation $E$ ($q_{1,0}$ to $q_{1,1}$) but then remains stable for 2 to 4 joins ($q_{1,2}$ to $q_{1,4}$) as $E$ is small compared to $|\mathcal{R}^*|$. Conflict query $q_2$ is slower as it has two subgoals, one of which is negative, which requires evaluation of nested disjunctions. The query for users $q_3$ is slowest as it includes a negative subgoal and ranges over the belief worlds of all users.

Table 2: Execution times and size of result sets for our seven example queries executed over a belief database with 10,000 annotations.

|  | $q_{1,0}$ | $q_{1,1}$ | $q_{1,2}$ | $q_{1,3}$ | $q_{1,4}$ | $q_2$ | $q_3$ |
|---|---|---|---|---|---|---|---|
| E(Time) [msec] | 105 | 145 | 146 | 152 | 144 | 436 | 4473 |
| $\sigma$(Time) [msec] | 120 | 168 | 153 | 162 | 162 | 186 | 661 |
| Result size | 1626 | 2816 | 2253 | 2061 | 1931 | 196 | 99 |

Overall, our experiments suggest that queries in a BDMS can be executed in reasonable amount of time (i.e. milliseconds) on top of a standard RDBMS.

### 6.3 Future Work

The dominant research challenge is to find techniques to decrease the *relative overhead* of belief databases. Recall that this overhead arises as result of the default assumption. For example, a generally accepted fact is, by default, believed by every user, and gets inserted into their respective belief worlds. More precisely, our current canonical Kripke structure stores $\bar{D}$, the set of all entailed beliefs, which means that it applies eagerly all instances of the default rule to $D$; this causes the database to increase. An alternative approach is to apply the default rule only selectively, or not at all, and to apply it only during query evaluation. This will complicate the query translation, but, at the same time, will drastically reduce the size of the database.

At the same time, a careful analysis and categorization of types of *queries that are common* in community databases will allow to optimize query time for certain queries. For example, conflict queries commonly focus on tuple-wide conflicts. Modeling functional dependencies between attributes during query translation will allow to back-chase tuple-wide attribute joins and reduce the number of necessary join attributes.

We are currently exploring these tradeoffs.

## 7. RELATED WORK

Work on *annotations management* in databases is often intertwined with provenance management [11] studying the propagation of annotations during query evaluation [12, 13]. In those contexts, annotations on data are commonly understood today as superimposed information that helps to explain, correct, or refute the base information [17, 36]. They are sometimes interpreted as colors, alternatively applied to individual values [7, 15], sets of values [23, 24] or as bundled tuples in tree fragments [11]. The important role that annotations play in science has been pointed out several times [4, 8, 10, 19, 20]. In all those settings, the semantic distinction between base information and annotations has remained blurred [10]. Annotations are simply additional data added to existing data [44]. In contrast, we propose to give a *concrete semantics to annotations* that helps users engage in a structured discussion on content and each other's annotations.

Several hardness results on inference in *modal logic* are well known [26, 31, 35]. Applications to reasoning about knowledge of agents in distributed settings is summarized in [21]. Modal logics have also been considered in databases before: Calvanese et al. [14] use the modal logic $K45_n^A$ to manage conflicts in a peer-to-peer system. Modalities are used to allow mappings between peers to exist even in the presence of conflicts. The work shares with us the common goal of using modalities to manage conflicting tuples, e.g. key violations. However, it differs as follows: (i) our modalities are part of the data. Users can add modalities to the data and ask queries with modalities to extract the desired facts from the database; (ii) in [14], the number of modalities is proportional to the size of the schema. In our case their number is proportional to the database; (iii) [14] considers only modalities of nesting depth 1. We allow arbitrary depth; and (iv) in [14] inference is in coNP. In contrast, ours is in PTIME (the difference comes from the fact that we restrict the use of negation to atomic tuples only). Another related work by Nguyen [38] constructs finite least Kripke models for different language fragments. The described algorithm runs in exponential time and returns a model with size $2^{O(n^3)}$, where $n$ is the size of input. In our work, we consider the fragment of *certain and impossible beliefs* and construct polynomial canonical representations. We also provide powerful insert, deletes, update functionalities to our model and can translate it into standard relations.

Another work that considers *key violations* is [22]. Here the approach differs from ours: key violations are allowed in the database, and are only resolved at query time through *repairs* [3]. Repairs are explored automatically by the system. At a high level, only those answers are returned that can be computed without any conflicts; there are no modalities, and hence the users have no control over conflict resolution.

There is a large body of work on managing uncertain and *incomplete information* [1, 2, 5, 16, 18, 28, 42, 47]. For example, Widom [48] describes the Christmas Bird Count as motivation, which is similar to our motivating scenario. Our work shares a similar motivation for information that is not certain. However, we do not measure, track, nor evaluate uncertainty; we rather allow conflicting user views and provide means for structured discourse inside a database.

We also share motivation with work on peer-data management and *collaborative data sharing* systems that have to deal with conflicting data and lack of consensus about which data is correct during integration [6, 29, 30, 34]. In contrast to these systems, we do not address the problem of schema integration. We consider conflicts at the data level within a given common schema. Systems like ORCHESTRA [27, 33, 45] enable different peers to accept or reject changes to shared data made by other peers. Each peer can have its own view of the data. This view, however, is materialized once for each peer in its separate database instance. In contrast, we propose to allow *conflicting information to co-exist in a single database* and we allow users to discuss these conflicts.

## 8. CONCLUSIONS

This paper describes a model of *database annotations* that allows users to annotate both content and other users' annotations with beliefs. It allows users to *collaboratively contribute and curate a shared data repository* as common today in large-scale scientific database applications. It also allows to explicitly *manage conflicts* and *inconsistencies* between different users and their views. We introduce the concept of *belief databases*, give a concrete application throughout the paper, show a polynomial-size encoding of our desired semantics on top of relational databases, and validate this concept with a prototype and tests on synthetic data.

## 9. ACKNOWLEDGEMENTS

We very much like to thank Karen Dvornich for valuable discussions and access to the data of the NatureMapping project, and the anonymous reviewers for detailed comments which greatly helped us improve the presentation of this paper. This research was supported in part by NSF under grants IIS-0513877, IIS-0713576, and CNS-0454425. Magda Balazinska was also supported by a Microsoft Research New Faculty Fellowship.

# APPENDIX

## A. NOMENCLATURE

| | |
|---|---|
| $\mathcal{R}$ | external schema with $\mathcal{R} = (R_1, \ldots, R_r)$ |
| $Tup$ | tuple universe with example tuple $t \in Tup$ |
| $I$ | standard database instance over the external schema |
| $W$ | belief world instance with $W = (I^+, I^-)$ |
| $[\![W]\!]$ | semantics of $W$ as incomplete database |
| $\Gamma$ | key constraints of a standard database instance $I$ |
| $\Gamma_1, \Gamma_2$ | consistency constraints for a belief world $W$ |
| $U$ | set of users $\{1, \ldots, m\}$ with $m = |U|$ |
| $\hat{U}^*$ | $\hat{U}^* = \{w \in U^* \mid w_{[i]} \neq w_{[i+1]}\}$ |
| $u, v, w$ | belief paths $\in \hat{U}^*$ |
| $\bar{w}$ | belief path of zero or more variables and constants |
| $\bar{x}$ | tuples of variables and constants |
| $d$ | nesting depth or belief path length with $d = |w|$ |
| $\varphi$ | belief statement $\varphi = \Box_w t^s$ with $s$ as sign |
| $D$ | data with belief statements $\{\varphi_1, \ldots, \varphi_n\}$, $n = |D|$ |
| $D_w$ | explicit belief world at $w$ (with belief path $w$) |
| $\bar{D}$ | entailment of $D$ |
| $\bar{D}_w$ | entailed belief world at $w$ |
| $g$ | number of subgoals of a belief conjunctive query |
| $K$ | Pointed Kripke structure with $K = (V, (W_v)_{v \in V}, (E_i)_{i \in U}, v_0)$ |
| $V$ | set of states |
| $W_v$ | belief world associated with a state $v \in V$ |
| $v_0$ | root of the Kripke structure; $v_0 \in V$ |
| $E_i$ | set of edges associated with each user $i \in U$ with $E_i \subseteq V \times V$ |
| $\mathcal{R}^*$ | internal schema; relational representation of a belief database |

## B. PROOFS

### B.1 Proof Proposition 5

PROPOSITION 5 (CONSISTENCY OF A BELIEF WORLD)
*A belief world $W$ is consistent iff it satisfies the following two constraints:*

$$\Gamma_1(W) \equiv \Gamma(I^+)$$
$$\Gamma_2(W) \equiv \forall t \in I^+ : t \notin I^-$$

PROOF. (1) We first show $\Gamma_1(W) \wedge \Gamma_2(W) \Rightarrow [\![W]\!] \neq \emptyset$: WLOG, we focus on a fixed key $k$. The procedure is implicitly assumed to be repeated for all $k \in \{key(t) \mid t \in Tup\}$. From $\Gamma_1$ follows that there is either (i) no or (ii) one positive tuple with key $k$ in the belief world $W$. From $\Gamma_2$ follows that there can be zero or more negative tuples with key $k$ in the belief world, but none that is already contained as positive tuple. In case (i), we can create a consistent complete database for key $k$ by making all tuples with this key $t.(key = k)$ negative. In case (ii), we make all tuples with the same key negative, except for the positive one.
(2) $[\![W]\!] \neq \emptyset \Rightarrow \Gamma_1(W) \wedge \Gamma_2(W)$: We again focus on a fixed key $k$. A complete database $I \in [\![W]\!]$ is consistent, if for each key $k$, there are either zero or one (positive) tuples in the database. Therefore $\Gamma_1$. A tuple cannot be at the same time be in the database and not be in the database. Therefore $\Gamma_2$. □

### B.2 Proof Proposition 7

PROPOSITION 7 (CERTAIN AND IMPOSSIBLE TUPLES)
*Let $t$ be a tuple in $Tup_i$, i.e. it is a typed tuple for relation $R_i$. Tuple $t$ is a positive belief for $W$ iff it is in $I^+$. It is a negative belief for $W$ iff it is either in $I^-$ ("stated negative") or if there is another tuple $t' \in I^+$ from $Tup_i$ with the same key ("unstated negative"):*

$$W \models t^+ \text{ iff } t \in I^+$$
$$W \models t^- \text{ iff } \underbrace{t \in I^-}_{stated} \vee \underbrace{\exists t' \in Tup_i . \big(t' \in I^+ \wedge t' \neq t \wedge key(t') = key(t)\big)}_{unstated}$$

PROOF. (1) What we have to show is $t \in I^+ \Leftrightarrow \nexists I . (I^+ \subseteq I \wedge I \cap I^- = \emptyset \wedge \Gamma(I) \wedge t \notin I)$. (1a) $\Rightarrow$: If $t \in I^+$ and $I^+ \subseteq I$, then $t \in I$. Hence, $\nexists I . (t \notin I)$. (1b) $\Leftarrow$: If $\forall I . (I^+ \subseteq I \wedge I \cap I^- = \emptyset \wedge \Gamma(I) \Rightarrow t \in I)$, then $t \in I^+$. As if $t \notin I^+$, then there is always some consistent $I . (\forall t' . (key(t') = key(t) \Rightarrow t' \notin I))$ for which $t \notin I$.
(2) $W \models t^-$ iff $t \in I^- \vee \exists t' \in I^+ . \big(key(t') = key(t) \wedge t' \neq t\big)$: From Def. 3 and Def. 6, we know that $W \models t^-$ iff $\forall I . (I^+ \subseteq I \wedge I \cap I^- = \emptyset \wedge \Gamma(I) \Rightarrow t \notin I)$. Hence, we have to show

$$t \in I^- \vee \exists t' \in I^+ . \big(key(t') = key(t) \wedge t' \neq t\big)$$
$$\Leftrightarrow \nexists I . (I^+ \subseteq I \wedge I \cap I^- = \emptyset \wedge \Gamma(I) \wedge t \in I)$$

(2a) $\Rightarrow$: If $t \in I^-$ then $\nexists I . (I \cap I^- = \emptyset \wedge \Gamma(I) \wedge t \in I)$. If $\exists t' \in I^+ . \big(key(t') = key(t) \wedge t' \neq t\big)$ then $\nexists I . (I^+ \subseteq I \wedge \Gamma(I) \wedge t \in I)$. (2b) $\Leftarrow$: $\forall I . (I^+ \subseteq I \wedge I \cap I^- = \emptyset \wedge \Gamma(I) \Rightarrow t \notin I)$ is equivalent to $\forall I . \big(I \cap I^- = \emptyset \Rightarrow t \notin I\big) \vee \forall I . \big(I^+ \subseteq I \wedge \Gamma(I) \Rightarrow t \notin I\big)$. The first proposition $\forall I . \big(I \cap I^- = \emptyset \Rightarrow t \notin I\big)$ is true iff $t \in I^-$. The second proposition $\forall I . \big(I^+ \subseteq I \wedge \Gamma(I) \Rightarrow t \notin I\big)$ is true iff $\exists t' \in I^+ . \big(key(t') = key(t) \wedge t' \neq t\big)$. □

### B.3 Proof Theorem 16

THEOREM 16 (CANONICAL KRIPKE STRUCTURE)
*(1) For any belief statement $\varphi$, $D \models \varphi$ iff $K(\bar{D}) \models \varphi$.*
*(2) $K(\bar{D})$ can be computed in time $O(m^d n)$, where $n$ is the size of the belief database $D$, $m$ the number of users and $d$ is the maximum length of any belief path in $D$.*

PROOF. (1) From Def. 12, $D \models \varphi \Leftrightarrow \varphi \in \bar{D}$. Hence, it suffices to show that $\bar{K}(D) \models \varphi \Leftrightarrow \varphi \in \bar{D}$. We proceed in 5 steps:

(1a) We first construct an empty infinite *Kripke tree frame* $T_K$ for $U$, i.e. a tree with root $v_0$ so that for each belief path $w \in \hat{U}^*$, there is exactly one node in $T_K$ whose path from the root is $w$. Therefore, for each node at depth $d \neq 0$ and incoming $i$-edge ($i \in U$), we create $m - 1$ child nodes at depth $d + 1$ for all $j \in U \setminus \{i\}$. This tree has 1 node at depth 0, $m$ nodes at depth 1, $m(m-1)$ nodes at depth 2, in general $m(m-1)^{d-1}$ nodes at depth $d$. The number $N$ of nodes with depth $\leq d$ is then $N(d) = 1 + m + m(m-1) + \ldots + m(m-1)^{d-1}$. Using the geometric series, we get $N(d) = 1 + \frac{m}{m-2}((m-1)^d - 1) = \mathcal{O}(m^d)$.

(1b) We next consider *Kripke trees*, i.e. Kripke tree frames with each node corresponding to a belief world with belief path corresponding to the path from the root to each node. We define a sequence $T_K(D^{(i)})$ of Kripke trees corresponding to the infinite sequence from Def. 9,

$$D^{(0)} = D$$
$$D^{(d+1)} = D^{(d)} \cup \{\Box_i \varphi \mid \varphi \in D^{(d)}, i \in U,$$
$$\text{belief path of } \Box_i \varphi \in \hat{U}^*,$$
$$D^{(k)} \cup \{\Box_i \varphi\} \text{ is consistent}\}$$

where for each $\varphi \in D^{(i)}$ with $\varphi = \Box_w t^s$, we add $t^s$ to the node with path $w$ in $T_K(D^{(i)})$. We call $T_K(\bar{D})$ the *canonical Kripke tree*. As there is exactly one node for each belief path $w \in \hat{U}^*$ with the path $w$ from the root, it follows that $(T_d(D^{(i)}), v_0) \models \varphi \Leftrightarrow T_K(D^{(i)}) \models \varphi$ if and only if $\varphi \in D^{(i)}$. In particular, $T_K(\bar{D}) \models \varphi \Leftrightarrow \varphi \in \bar{D}$. Figure 7 shows the first two Kripke trees up to depth 2 for our running example $D$ from Fig. 2.

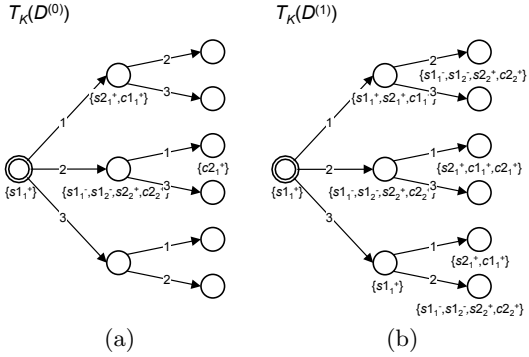

(a)      (b)

**Figure 7: Sequence of Kripke trees $T_K^{(i)}$ cut off at depth 2 for our running example.**

(1c) We next show that subtrees of $T_K(\bar{D})$ that do no contain $States(D)$ can be pruned and replaced by an appropriate back edge, so that for the resulting model $T'_K(\bar{D})$, it holds $T'_K(\bar{D}) \models \varphi \Leftrightarrow T_K(\bar{D}) \models \varphi$: Consider the general subtree starting from node #2 with path $w = i \cdot v \cdot j$ ($v, w \in \hat{U}^*; i, j \in \{\varepsilon\} \cup U$) in Fig. 8a. Assume the subtree contains no $States(D)$, i.e. there is no $\varphi \in D$ with belief path $w \cdot u$ ($w \cdot u \in \hat{U}^*$). Then the subtree starting from node #2 is isomorph to the subtree starting at node #3 with path $v \cdot j$ which we show as follows:

 (i) each node at depth $d = |w \cdot u|$ in subtree #2 can be mapped to a node at depth $d - 1$ in subtree at #3 in such a way that the edges of #2 map to edges in #3. This follows inductively by starting to map node #2 to #3 and repeating at each subsequent depth;

 (ii) each belief world at a node in #2 is the same as the corresponding belief world at #3. This follows from the fact that for each node with path $w \cdot u$ in #2, $D_{w \cdot u} = \{\}$. Hence each tuple in subtree #2 of $T_K(\bar{D})$ is inserted by the default rule from Def. 9 which inserts each tuple $t^s$ in $\varphi = \Box_{v \cdot j \cdot u} t^s$ from the node with path $v \cdot j \cdot u$ in #3 into the corresponding node with path $i \cdot v \cdot j \cdot u$ in #2. Hence, corresponding nodes have the same belief worlds.

As a consequence, we can create a new Kripke tree $T'_K(\bar{D})$ from $T_K(\bar{D})$ with pruned subtree #2 and replaced forward $j$-edge (#1, #2) by a back $j$-edge (#1, #3) as shown in Fig. 8b. We know $(T_K(\bar{D}), \#2) \models \varphi \Leftrightarrow (T_K(\bar{D}), \#3) \models \varphi$. It follows: $T_K(\bar{D}) \models \Box_{i \cdot v \cdot j} \varphi \Leftrightarrow (T_K(\bar{D}), v_0) \models \Box_{i \cdot v \cdot j} \varphi \Leftrightarrow (T_K(\bar{D}), \#2) \models \varphi \Leftrightarrow (T_K(\bar{D}), \#3) \models \varphi \Leftrightarrow (T'_K(\bar{D}), v_0) \models \varphi \Leftrightarrow T'_K(\bar{D}) \models \Box_{i \cdot v \cdot j} \varphi$. Hence, the original and the pruned Kripke tree have the same theory: $T_K(\bar{D}) \models \varphi \Leftrightarrow T'_K(\bar{D}) \models \varphi$. Note that node #3 with path $v \cdot j$ is the node with the largest suffix of node #2 with path $i \cdot v \cdot j$.

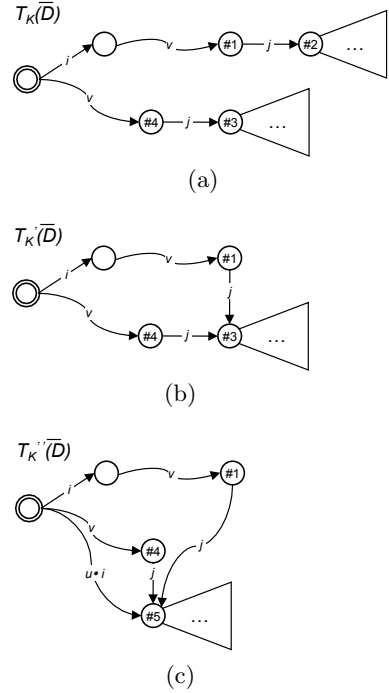

(a)

(b)

(c)

**Figure 8: The pruned Kripke trees $T'_K(\bar{D})$ and $T''_K(\bar{D})$ have the same theory as $T_K(\bar{D})$.**

(1d) If the node $\#3 \notin States(D)$, then #3 can itself be pruned according to (1c), i.e. we can replace the $j$-edge (#4, #3) by a $j$-edge from #4 to the largest suffix of #3, say #5. Since the subtrees #3 and #5 are ismorphic, so are #2 and #3. Hence, we can further replace the $j$-edge (#1, #3) by $j$-edge (#1, #5), and we still have $T_K(\bar{D}) \models \varphi \Leftrightarrow T''_K(\bar{D}) \models \varphi$. As a consequence, each forward $j$-edge (#1, #2) with $\#2 \notin States(D)$ can be replaced with a back $j$-edge (#1, #5), where #5 has the largest suffix path of #1 and $\in States(D)$.

If we repeat this pruning for each edge between a node $\#1 \in States(D)$ and a node $\#2 \notin States(D)$, then we get exactly the construction of the canoni-cal Kripke model $K(\bar{D})$ in Def. 16. Hence, $T_K(\bar{D}) \models \varphi \Leftrightarrow K(\bar{D}) \models \varphi$, and $\varphi \in \bar{D} \Leftrightarrow K(\bar{D}) \models \varphi$.

(2) We first give an alternative construction of $K(\bar{D})$ that avoids the intermediate infinite Kripke tree and then evaluate the complexity of this method.

(2a) Fix a node in the canonical Kripke tree $T_K(\bar{D})$ with path $w = u \cdot v \cdot x \cdot y$ $(u, v, x, y, w \in \hat{U}^*)$ and $|w| = d$. For a tuple $t^s$ to be in the node with path $w$, either (i) $t^s \in D_w$ or (ii) $\exists w' = u \cdot v$ with $t^s \in D_{w'}$ and $\forall w'' = u \cdot v \cdot x : D_{w''} \cup t^s$ is consistent. It follows that the content of a node with path $w$ and $k = |w|$ in the canonical Kripke tree $T_K(\bar{D})$ (and hence, the world $\bar{D}_w$) is the same as in the Kripke tree in the canonical Kripke tree $T_K(D^{(d)})$ (and hence, the world $D^{(d)}_w$), and it can be deduced by only analyzing all suffix worlds of $w$: $\{D_w, D_{w_{[2,d]}}, \ldots D_{w_{[d]}}, D_\varepsilon\}$. Figure 9 illustrates that the content of world $D^{(d)}_w$ in the left lower corner, and hence $\bar{D}_w$, can be deduced in the following iterative way: start with the root world $D_\varepsilon$. Insert all tuples from $\varepsilon$ into the belief world $D_{[d]}$ which are consistent wit $D_{[d]}$. Repeat this step for all belief worlds until $D_w$.

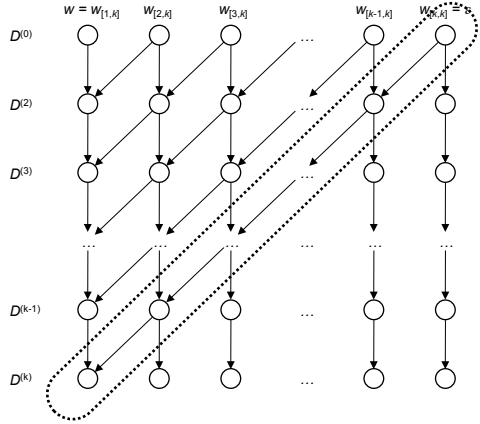

**Figure 9:** Calculating $\bar{D}_w$ can be done in $d = |w|$ steps by consecutively analyzing all suffix states of $w$ with increasing depth and inserting all tuples from $D_{[x,d]}$ into $D_{[x-1,d]}$ which are consistent with the tuples already in $D_{[x-1,d]}$.

(2b) The algorithm for constructing the whole canonical Kripke model $K(\bar{D})$ starts with (i) the *canonical Kripke frame*, i.e. all nodes that correspond to the $States(D)$, forward edges; (ii) it then inserts all back edges to the largest suffixes according to Def. 16. (iii) It then determines for each node in the model its largest suffix node in order to constructs an *inverted suffix tree*, i.e. the tree from the root node to all other nodes where a node #1 with path $i \cdot v$ is a child of another node #4 with path $v$ if the path of #4 is a suffix of the path of #1 and no deeper node is a suffix of #1. (iv) Use either a breath-first algorithm that calculates for each state the overriding union with its largest suffix state, or a depth-first algorithm that traverses an inverted suffix tree and calculates the overriding union at each step.

(2c) Complexity:
  (i) Construction of the *canonical Kripke frame* (i.e. the canonical Kripke structure without any tuples) without the back edges takes $\mathcal{O}(nd)$, as for each of $n$ belief statements $\varphi \in D$ with belief path $w$ we need $|w| \leq d$ operations.
  (ii) Construction of the back edges takes $\mathcal{O}(m^{d+1}d^2)$ time: (a) a canonical Kripke frame with all worlds $w \in \hat{U}^*$ and $|w| \leq d$ has $N(d) = 1 + \frac{m}{m-2}((m-1)^d - 1) = \mathcal{O}(m^d)$ nodes. The root node has $m$, each other node $m-1$ outgoing edges. The number of edges is hence $E(d) = m + \frac{m(m-1)}{m-2}((m-1)^d - 1) = \mathcal{O}(m^{d+1})$. The number of leaf nodes is $m(m-1)^{d-1} = \mathcal{O}(m^d)$; it hence needs $m(m-1)^d = \mathcal{O}(m^{d+1})$ back edges. For each such edge, we have to find the largest suffix node, which takes $\frac{d(d-1)}{2} = \mathcal{O}(d^2)$ in the worst case and with a naive algorithm.
  (iii) For each of the $\mathcal{O}(m^d)$ nodes in the Kripke frame, determine the node in the model with the largest suffix of its path. This can be bound by $\mathcal{O}(m^d(d-1))$ analog to (ii) above.
  (iv) Inserting all implicit beliefs is $\mathcal{O}(m^d n)$. For that, assume the worst case of a canonical Kripke frame with all worlds $w \in \hat{U}^*$ and $|w| \leq d$. Insert of a tuple at the root leads to checking for all $\mathcal{O}(m^d)$ nodes in the model. The insert/check at each node can be performed in $\mathcal{O}(1)$ with a hash index on the key attribute. Hence, $\mathcal{O}(m^d n)$.
  (v) Note that $n \geq m(m-1)^{d-1}$ for the bound in (ii) as minimum number of annotations for the canonical Kripke model to include all states with $w \in \hat{U}^*$ and $|w| \leq d$. Further note that $(m-1)^{d-1} \geq d^2$ for $(m \geq 3, d \geq 7)$ or $(m \geq 4, d \geq 3)$ or $m \geq 5$. Hence, we can bound $\mathcal{O}(m^{d+1}d^2)$ by the looser bound $\mathcal{O}(m^d n)$, which is the overall bound.

□

## C. DEFAULT LOGIC

We shortly review Reiter's default logic [41] before drawing the connection to our message board assumption. We mostly follow the exposition and notation of Gottlob [26] and Brachman and Levesque [9].

### C.1 Default logic primer

A *default rule* $d$ is a configuration of the form

$$\frac{\alpha : \beta}{\omega},$$

where $\alpha$, $\beta$, and $\omega$ are propositional sentences. Usually, $\alpha$ is called the *prerequisite*, $\beta$ the *justification*, and $\omega$ the *consequence* of $d$. A default rule $d$ is satisfied by a deductively closed set of sentences $S$ if, whenever $\alpha$ is an element of $S$ and $\beta$ is consistent with $S$, then $\omega$ also is an element of $S$. A *normal default rule* is one where justification and consequence are the same:
$$\frac{\alpha : \omega}{\omega} \ .$$

A default that contains formulae with free variables is sometimes called a *default schema*. It defines the set of all default rules obtained for all ground substitutions that assign values to all free variables occurring in the schema. A propositional *default theory* is a pair $T = (W, D)$, where $W$ is a finite set of propositional sentences, sometimes called the *background theory*, and $D$ is a finite set of default rules.

Informally, an *extension* $E$ of a default theory $(W, D)$ is a grounded minimal deductively closed set of propositional formulae containing $W$ and satisfying all defaults of $D$. Hence, extensions are (minimal) fixed points of the operator $D$, namely, that further application of the default rules in $D$ to the sentences in an extension has no effect. More formally: Let $(W, D)$ be a default theory. For any set $S$ of propositional formulae, let $\Gamma(S)$ be the smallest set $U$ satisfying the following three properties:
(1) $W \subseteq U$.
(2) $U$ is deductively closed.
(3) If $\frac{\alpha:\beta}{\omega} \in D$ and $\alpha \in U$ and $\neg\beta \notin S$, then $\omega \in U$.
An extension of $(W, D)$ is a fixpoint of $\Gamma$, i.e. a set $E$ of propositional formulae satisfying $\Gamma(E) = E$.

An alternative definition is as follows: Given a default theory $T = (W, D)$. A set of sentences $E$ is an extension of $T$ if and only if for every sentence $\varphi$,

$$\varphi \in E \text{ iff } W \cup \{\omega | \tfrac{\alpha:\beta}{\omega} \in D, \alpha \in E, \neg\beta \notin E\} \models \varphi$$

An equivalent algorithmic definition is as follows. A default $\frac{\alpha:\beta}{\omega}$ is applicable to a propositional theory $W$ if $W \models \alpha$ and $W \cup \beta$ is consistent. The application of this default to $W$ leads to the theory $W \cup \omega$.

A default theory can have one, more or no extension. A *normal default theory*, i.e. a default theory that has only normal default rules, has at least one extension.

## C.2 Default beliefs for belief databases

We can define the belief theory $T$ of a belief database $D$ of a finite set of belief statements as the pair $(D, A)$, where $A$ is the set of default assumptions consisting of one default schema, the message board assumption

$$\frac{\varphi : \Box_i \varphi}{\Box_i \varphi} \ .$$

In our notation, the extension $\bar{D}$ of a belief database consists of the explicitly annotated belief statements $D$ and a set of implied belief statements from the default assumption, such that no additional beliefs can be implied from $\bar{D}$ that are are consistent with $\bar{D}$. We call all $\varphi \in D$ the *explicit* belief statements, and all $\varphi \in \bar{D} \setminus D$ the *implicit* belief statements of a belief database.

Replacing Def. 9 and Def. 10 with the following definition gives an alternative definition of the semantics of a belief database together with Def. 12:

DEFINITION 19 (EXTENSION OF A BELIEF DATABASE). *Given a belief database $D$ and a set of default assumptions $A = \{d_s\}$ consisting of one normal default schema*

$$d_s \equiv \frac{\varphi : \Box_i \varphi}{\Box_i \varphi} \ ,$$

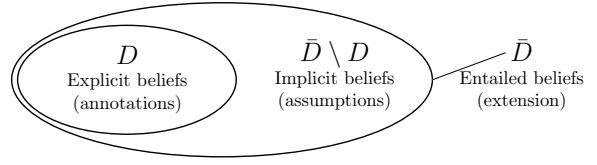

Figure 10: A belief database "contains" or entails more than just the explicit belief annotations.

*where $\varphi$ and $\Box_i \varphi$ are belief statements over the external schema. A set of sentences $\bar{D}$ is an extension of a belief database if and only if for every sentence $\varphi$,*

$$\varphi \in \bar{D} \text{ iff } \varphi \in D \vee \varphi \in \{\omega \mid \tfrac{\alpha:\beta}{\omega} \in A, \alpha \in \bar{D}, \beta \text{ is consistent with } \bar{D}\}$$

The important consequence of the following lemma is that the order in which default rules from the default schema are applied does not matter and we have one unique stable model for $D$ (observation in Sect. 3.4). This observation allows an efficient depth-first construction of the materialized canonical belief database.

LEMMA 20. *If $D$ is consistent, then $D$ has exactly one consistent extension $\bar{D}$.*

PROOF. (1) There is at least one consistent extension: An extension $E$ of a default theory is inconsistent if and only if the background theory is inconsistent and every default theory has at least one extension. As our default theory is normal, we have at least one consistent extension.

(2) There is maximal one consistent extension: Assume there exist two consistent extensions $\bar{D}$ and $\bar{D}'$. WLOG, there must then be one belief statement $\varphi \in \bar{D}'$ that is not in $\bar{D}$. Let $\varphi = \Box_w t^s$ with $w = v \cdot i$. As $D \subseteq \bar{D}$ and $D \subseteq \bar{D}'$, $\varphi$ must be implicit and there must be a grounded default rule

$$d_1 \equiv \frac{\Box_v t^s : \Box_{v \cdot i} t^s}{\Box_{v \cdot i} t^s} \ ,$$

which is satisfied for $\bar{D}'$, but not for $\bar{D}$. This can happen either because (i) the prerequisite $\Box_v t^s \in \bar{D}'$, but $\notin \bar{D}$; or (ii) the justification $\Box_{v \cdot i} t^s$ is consistent with $\bar{D}'$, but not with $\bar{D}$. We only have to focus on case (ii) as case (i) can be reduced to at least one occurrence of case (ii) by backchaining. So it suffices to disprove (ii).

So assume that $\Box_v t^s \in \bar{D}$ and $\in \bar{D}'$, but $\Box_{v \cdot i} t^s$ is consistent with $\bar{D}'$ and inconsistent with $\bar{D}$. For that to happen, there must be a belief statement $\Box_{v \cdot i} t'^{s'} \in \bar{D}$ but $\notin \bar{D}'$, which is inconsistent with $\Box_{v \cdot i} t^s$ and, hence, the grounded tuples $t^s$ and $t'^{s'}$ do not fulfill the requirements $\Gamma_1$ and $\Gamma_2$ of Prop. 5 for $\bar{D}'_w$ to be consistent. This necessarily implicit belief statement can only be in $\bar{D}$ because of another grounded default rule

$$d_2 \equiv \frac{\Box_v t'^{s'} : \Box_{v \cdot i} t'^{s'}}{\Box_{v \cdot i} t'^{s'}} \ .$$

For the prerequisite of $d_2$ to be satisfied, $\Box_v t'^{s'}$ has to be in $\bar{D}$. Hence, the $\Box_v t^s$ and $\Box_v t'^{s'}$ would have to be in $\bar{D}$, hence the belief world $\bar{D}_v$ is inconsistent which contradicts our assumptions. □

**Some thoughts.** (1) In default logic, the *extension* of a logical theory $W$ creates a new logical theory $E$ that "extends" $W$. In contrast, in standard database nomenclature, the *extensional* database refers to the explicitly stored tables and *intensional* to the relations defined or implied by rules. To avoid a possible naming ambiguity, we call *explicit* all belief statements $\varphi \in D$ and *implicit* all $\varphi \in \bar{D} \setminus D$. (2) In default logic, $D$ stands for the set of default rules, whereas we use $D$ for the explicit part of a belief database corresponding to the standard usage in database literature. (3) Our default schema $d_s$ defines infinitely many default rules and an infinite extension. (4) Note that consistency is defined by extended key constraints and differs from the propositional case: $\varphi \cup \bar{D}$ consistent $\not\Leftrightarrow \neg\varphi \notin \bar{D}$.

## C.3 Errata

This report includes the following corrections over the final PVLDB version:

- Sect. 5.1: Relation S:
  $$S(wid(w), wid(dss(w_{[2,d]})))$$
  instead of
  $$S(wid(w), wid(dss(w)))$$

- Sect. 5.3: 3rd paragraph:
  "Given the world id $y$ of $w$ ..."
  instead of
  "Given the world id $x$ of $w$ ..."

- Sect. 5.3: Algorithm 2:

  7 For all worlds $v \cdot w_{[1,d-1]}$ for which $w$ is the deepest suffix state for $v \cdot w$, update the $w_{[d]}$-edge:
  8 $\triangle S(x, \mathsf{dss}(w_{[2,d]}))$

  instead of

  7 For all worlds $v \cdot w_{[1,d-1]}$ for which $w$ is the deepest suffix state, update the $w_d$-edge:
  8 $\triangle S(x, \mathsf{dss}(w))$

- Sect. 5.3: Algorithm 3:

  1 **for** $p = 1 \ldots (d+1)$ **do**
  $\quad T(z,y)\text{:--}E^*(0, w_{[p,d]}, z), D(z,y)$

  instead of

  1 **for** $p = 2 \ldots (d+1)$ **do**
  $\quad T(z,y)\text{:--}E^*(0, w_{[p,d]}, x), D(z,y)$